\documentstyle[12pt]{article}

\begin{document}
\begin{center}
{\large\bf Non-Parallel Electric and Magnetic Fields in a
Gravitational Background, Stationary Gravitational Waves and Gravitons}

\vspace{0.2cm}
{\sl Carlos Pinheiro}$^{\ast}$

\footnotesize{Departamento de F\'{\i}sica, CCE\\[-2mm]
Universidade Federal do Esp\'{\i}rito Santo -- UFES\\[-2mm]
Av. Fernando Ferrari S/N, Campus Goiabeira\\[-2mm]
29060-900 Vit\'oria, ES -- Brazil \\[-2mm]
e-mail: fcpnunes\@@cce.ufes.br/maria\@@gbl.com.br}

\vspace{0.1cm}
\normalsize
{\sl J.A. Helay\"el-Neto$^{\dag}$}

\footnotesize{Centro Brasileiro de Pesquisas F\'{\i}sicas  - CBPF\\[-2mm]
Rua Dr. Xavier Sigaud, 150, 22290-180 Rio de Janeiro, RJ -- Brazil\\[-2mm]
and\\[-2mm]
Universidade Cat\'olica de Petr\'opolis, UCP}

\vspace{0.1cm}
\normalsize
{\sl Gilmar S. Dias$^{\ddag}$}

\footnotesize{Departamento de F\'{\i}sica, CCE, 
\\[-2mm]
Universidade Federal do Esp\'{\i}rito Santo -- UFES\\[-2mm]
Av. Fernando Ferrari S/N, Campus Goiabeira,\\[-2mm]
29060-900  Vit\'oria, ES -- Brazil \\[-2mm]
and\\[-2mm]
Escola T\'ecnica Federal ETFES}

\vspace{0.1cm}
{\sl F.C. Khanna}\\
\footnotesize{Theoretical Physics Institute, Dept. of Physics\\
University of Alberta,\\
Edmonton, AB T6G2J1, Canada\\
and\\
TRIUMF, 4004, Wesbrook Mall,\\
V6T2A3, Vancouver, BC, Canada.
khanna@@phys.ualberta.ca}
\end{center}

\vspace{1mm}
\begin{abstract}
The existence of an electromagnetic field with parallel electric and
magnetic field components in the presence of a gravitational
field is considered. A non-parallel solution is shown to exist. Next,
we analyse 
the possibility of finding stationary gravitational waves in nature.
Finally, we construct a $D=4$ effective quantum gravity model.
Tree-level unitarity is verified.
\end{abstract}

\vspace{2mm}
PACS: 1225 Models for Gravitational Interactions\\[-2mm]

\newpage
\setcounter{equation}{0}
\section{Electric and Magnetic Field in a Gravitational Background}
\paragraph*{}

Based on a series of papers by Brownstein [1] and Salingaros [2], we
consider here the possibility of the existence of an eletromagnetic
field whose electric and magnetic field components are parallel in
the presence of a gravitational field. The coupling between the
electromagnetic sector and the gravitational backgrounds is
accomplished by means of the action
\begin{equation} 
S = \int \sqrt{-\tilde{g}} \ \left(-\frac{1}{4} \ F_{\mu\nu}
F^{\mu\nu}\right) d^4x \ ,
\end{equation}
where
\[
\tilde{g} = det \ (g_{\mu\nu}) \ ,
\]
and
\[
F_{\mu\nu} = \partial_{\mu}A_{\nu} - \partial_{\nu}A_{\mu} \ .
\]
>From the above action, the following field-equations follow:
\begin{eqnarray}
&&{\cal D}_{\mu}F^{\mu\nu} = \partial_{\mu}F^{\mu\nu} +
\Gamma^{\beta}_{\beta\lambda}F^{\lambda\nu} +
\Gamma^{\nu}_{\mu\lambda} F^{\mu\lambda} = J^{\nu} \ , \\
&& \nonumber \\
&& {\cal D}_{\mu}F_{\nu\beta} + {\cal D}_{\nu}F_{\beta\mu} + {\cal
D}_{\beta}F_{\mu\nu} = 0 \ .
\end{eqnarray}
Choosing the background to be described by the F.R.W metric,
\begin{equation}
dS^2 = dt^2-a^2 (t) \left[\frac{dr^2}{1-Ar^2} + r^2 d\theta^2 
+ r^2 sin^2\theta d\phi^2\right]
\ ,
\end{equation}
the Maxwell equations in the absence of electromagnetic sources are 
\begin{eqnarray}
&& \vec{\nabla}\cdot \vec{E} = g\vec{\nabla} f\cdot \vec{E} \ ,\\
&& \vec{\nabla}\cdot \vec{B} = 0 \ , \nonumber \\
&& \vec{\nabla} \times \vec{E} = -\frac{\partial
\vec{B}}{\partial t} \ , \nonumber \\
&& \vec{\nabla}\times \vec{B} = \frac{\partial\vec{E}}{\partial t} - g \
\frac{\partial f}{\partial t} \ \vec{E} + g\vec{\nabla} f\times
\vec{B} - \Gamma^i_{\mu\beta}F^{\mu\beta} \ , \nonumber 
\end{eqnarray}
where 
\begin{equation}
g = \frac{\sqrt{1-Ar^2}}{a^3r^2sin\theta} \ , \hspace{2cm} f =
\frac{a^3r^2sin\theta}{\sqrt{1-Ar^2}} \ ,
\end{equation}
and $A=+1,0,-1$. \\
The wave-equations for $\vec{E}$ and $\vec{B}$ are found to be:
\begin{eqnarray}
&& \nabla^2\vec{E} - \frac{\partial^2\vec{E}}{\partial t^2} =
\vec{\nabla} (g\vec{\nabla} f\cdot\vec{E}) - \ \frac{\partial
g}{\partial t} \ \frac{\partial f}{\partial t} \vec{E} - g \
\frac{\partial^2 f}{\partial t^2} \vec{E} -   g \
\frac{\partial f}{\partial t}\ \frac{\partial\vec{E}}{\partial t} + \\
&& \nonumber \\
&& \frac{\partial g}{\partial t}\vec{\nabla} f \times \vec{B} + g \ 
\frac{\partial}{\partial t}\vec{\nabla} f \times \vec{B} + g\vec{\nabla} f \times 
\frac{\partial\vec{B}}{\partial t} - \frac{\partial}{\partial t}
(\Gamma^i_{\mu\nu}F^{\mu\beta})\ , \nonumber \\
&& \nonumber \\
&& \nabla^2\vec{B} - \frac{\partial^2\vec{B}}{\partial t^2} =
\vec{\nabla}\times \left(g \ \frac{\partial f}{\partial t} \vec{E} -
g\vec{\nabla} f \times \vec{B} + \Gamma^i_{\mu\beta}
F^{\mu\beta}\right)\ .
\end{eqnarray}

Now, due to the  presence of the gravitational background, we
have explicitly built up a solution for $\vec{E}$ and $\vec{B}$ that
is different from the one obtained
by Brownstein [1] and Salingaros [2]. These authors state that it
is always possible to find solutions for parallel $\vec{E}$ and
$\vec{B}$ in plasma physics or in an  astrophysics plasma. However,
contrary to their result, we have found non-parallel solutions due to
the non-flat background of gravity:
\begin{eqnarray}
&& \vec{E} = \hat{i} \Big(\sin \theta G(r,t,\theta ) - \cos \theta
F(r,t)\Big)ka \cos (kz) \cos (\omega t) +  \\
&& +\vec{j} \Big(\cos \theta F(r,t) - \sin \theta G(r,t,\theta ) ka \
\sin (kz) \cos (\omega z)\Big)  + \nonumber \\
&& + \vec{k} \Big[\Big(sin\ \theta \ \cos\ \varphi\ F(r,t) + \cos \ \theta \
\cos \ \varphi \ G(r,t,\theta )\Big)ka
\ \cos (kz) \cos (\omega t) + \nonumber \\
&& - \Big(\sin\ \theta \ \sin \ \varphi F(r,t) + \cos \theta \ sin \
\varphi \
G_{(r,t,\theta )} ka \ \sin (kz) \cos (\omega t)\Big) \Big] \nonumber 
\end{eqnarray} 
and
\begin{equation}
\vec{B} = ka\big[\hat{i} \ \sin (kz) + \hat{j} \cos (kz) \big] \cos (\omega t)
\end{equation}
where the functions $G(r,t,\theta ) =
\displaystyle{\frac{a\cot g\theta}{3\dot{a}r}}$   and 
\begin{equation}
F(r,t) = \frac{2a}{3\dot{a}r} + \frac{Aar}{3\dot{a}(1-Ar^2)}
\end{equation}
are the metric contribution.

\newpage
\section{Stationary Gravitational Waves and Gravitons}\setcounter{equation}{0}
\paragraph*{}

Now, we analyse the possibility of finding stationary gravitational
waves. From a phenomenological viewpoint, a distribution of black
holes could play  the role of knots for the non-propagating
gravitational waves. We postulate the equation that may lead to this
sort of waves to be of the form
\begin{eqnarray}
&&R_{\mu\nu} = \kappa \Lambda h_{\mu\nu} \ , \\
&&\nonumber \\
&& g_{\mu\nu} (x) = \eta_{\mu\nu} + \kappa h_{\mu\nu} \ ,
\end{eqnarray}
where $\Lambda$ is the cosmological constant. These equations yield:
\begin{equation}
\partial_{\beta}\partial_{\nu}h^{\beta}_{\mu} +
\partial_{\beta}\partial_{\mu}h^{\beta}_{\nu} - \Box h_{\mu\nu} -
\partial_{\mu}\partial_{\nu} h^{\beta}_{\beta} = \Lambda h_{\mu\nu} \
.
\end{equation}
Now, solutions of the form
\begin{equation}
h_{\mu\nu} = C_{\mu\nu} (z) f (t) \ ,
\end{equation}

\begin{equation}
h_{\mu\nu}=\left(
\begin{array}{llll}
A_{00} & 0 & 0 & 0 \\
0 & A_{11} & A_{12} & 0 \\
0 & A_{12} & -A_{11} & 0 \\
0 & 0 & 0 & A_{00}
\end{array}
\right)e^{i\tilde{k}z} cos \ \omega t\ ,
\end{equation}
can be found, where $A_{00}$, $A_{11}$ and $A_{12}$ are free
parameters, whereas $\tilde{k}=\sqrt{\Lambda -\omega^2}$ is the wave
number. Having in mind that $\Lambda$ is a small number, the
frequency $\omega$ must be extremely small. This forces us to search
for a mechanism to detect such low-frequency stationary waves.

The equations of motion are derived from the Lagrangian density
\begin{equation}
{\cal L}_H = \frac{1}{2} H^{\mu\nu}\Box H_{\mu\nu} - \frac{1}{4} H\Box H -
\frac{1}{2} \ H^{\mu\nu}\partial_{\mu}\partial_{\alpha}
H^{\alpha}_{\nu} - \frac{1}{2} \ H^{\mu\nu}
\partial_{\nu}\partial_{\alpha} H^{\alpha}_{\mu} - \frac{1}{2}\ \Lambda H^{\mu\nu}
H_{\mu\nu} + \frac{1}{4} \ \Lambda H^2\ ,
\end{equation}  
where
\[
H^{\alpha}_{\nu} = h^{\alpha}_{\nu} -\frac{1}{2} \
\delta^{\alpha}_{\nu} h
\]
and the bilinear form operator of lagrangian (2.17) is given by
\begin{eqnarray}
&&\Theta_{\mu\nu ,\kappa\lambda} = (\Box -\Lambda ) P^{(2)} - \Lambda
P_m^{(1)} + \frac{5}{2} (\Box -\Lambda ) P_s^{(0)} - 
\frac{(\Lambda +3\Box )}{2} \ P_w^{(0)} \nonumber \\
&& + \frac{\sqrt{3}}{2} (\Lambda
-\Box ) \ P_{sw}^{(0)} + \frac{\sqrt{3}}{2} \ (\Lambda -\Box ) \
P_{ws}^{(0)} \ ,
\end{eqnarray}
and $P^{(i)}$, $i=0,1,2$, are spin projection operators in the space
of rank-2 symmetric tensors. The graviton propagator is given by:
\begin{equation}
\langle T(h_{\mu\nu}(x);h_{\kappa\lambda}(y))\rangle =
i\Theta^{-1}_{\mu\nu ,\kappa\lambda} \delta^4 (x-y) 
\end{equation}
where
\begin{equation}
\Theta^{-1} = [XP^{(2)} + YP_m^{(1)} + ZP_s^{(0)} + WP_w^{(0)} +
RP^{(0)}_{sw} + SP^{(0)}_{ws} ]_{\mu\nu ,\kappa\lambda}
\end{equation}
with
\begin{equation}
X = -\ \frac{1}{\Lambda -\Box}\ , \qquad Y=-\ \frac{1}{\Lambda}\ , \qquad
Z=-\ \frac{\Lambda +3\Box}{\Lambda^2+8\Lambda\Box - 9\Box^2}\  ,
\end{equation}

\[
W = -\ \frac{5}{\Lambda -9\Box}\ , \qquad R=-\ \frac{\sqrt{3}}{\Lambda
+9\Box}\qquad \mbox{and}\ \qquad S =-\ \frac{\sqrt{3}}{\Lambda
+9\Box}\ .
\]
>From this propagator,  a current-current amplitude is obtained and 
the tree-level unitarity [3] is discussed. Three massive excitations
are found: They are a spin-2 quantum with mass equal to $k^2=\Lambda$
and two massive spin-0 quanta with masses equal to $k^2=\Lambda$ and
$k^2=-\ \displaystyle{\frac{1}{9}\Lambda}$. The spin-2 is a physical
one: the imaginary part of the residue of the amplitude at the pole
$k^2=\Lambda$ is positive, so that it does not lead to a ghost. It
remains to be shown that the tachyonic pole,
$k^2=\displaystyle{-\ \frac{1}{9}\Lambda}$, is non-dynamical or
decouples through some constraint on the sources. 

We conclude, then, that in a gravitational  background it is always
possible to find non-parallel electric and magnetic fields. It is
the gravitational field that breaks the parallel configuration of
$\vec{E}$ and $\vec{B}$ [1,2]. Furthermore, a stationary gravitational wave
equation is postulated and a particular solution is found. We argue
that such a solution  is likely to be found in Black Hole distributions.
Finally, we set up an effective quantum gravity model where the
necessary condition for the tree-level unitarity for the spin-2
sector is respected. The model is infrared finite though
non-renormalizable in the ultraviolet limit.

\section*{Acknowledgements}
\paragraph*{}

The authors acknowledge Dr. Berth Sch\"oer for helpful suggestions
and technical discussions. Thanks are also due to Gentil O. Pires and
Manoelito M. de Souza for a critical reading of the manuscript. We
would like also to thank the Department of Physics, University of
Alberta for their hospitality and Dr. Don N. Page for his kindness and attention
with  me at Univertsity of Alberta. 
This work was partially supported by Conselho Nacional de
Desenvolvimento Cient\'{\i}fico e Tecnol\'ogico, CNPq, Brazil.

\end{document}